\documentclass[12pt]{article}
\def\beq {\begin{eqnarray}}
\def\eeq {\end{eqnarray}}
\def\be {\begin{equation}}
\def\ee {\end{equation}}
\def \half {{\textstyle{1\over 2}}}
\def\Tr {{\rm Tr}}
\def\dag {\dagger}
\def\del {\partial}

\def\a {\alpha}
\def\b {\beta}
\def\g {{\gamma}}
\def\d {{\delta}}
\def\e {\epsilon}

\def \bz {\bar{z}}

\def\ra {\rangle}
\def\la {\langle}

\begin{document}

\begin{titlepage}

\pagestyle{empty}
\begin{center}
\rightline{CCNY-HEP-02/03}

\vspace{1.0truein} {\Large \bf  Quantum Hall Effect in Higher Dimensions}\\
\vspace{.5in}  { DIMITRA KARABALI
$^{a,c}$ and  V.P. NAIR $^{b,c}$ \footnote{{\it e-mail addresses:}
karabali@alpha.lehman.cuny.edu, vpn@sci.ccny.cuny.edu}
}\\
\vspace{.3in}  {\it $^a$ Department of Physics and Astronomy,
Lehman College of the CUNY\\ Bronx, NY 10468}\\
\vspace {.1in}
{\it $^b$ Physics Department, City College of the CUNY\\  New
York, NY 10031}\\
\vspace {.1in} {\it $^c$ The Graduate School and
University Center, CUNY\\ New York, NY 10016}\\

\end{center}
\vspace{1.5in}

\centerline{\bf Abstract}
\vskip .1in
Following recent work on the quantum Hall effect on
$S^4$, we solve the Landau problem on
the complex projective spaces
${\bf C}P^k$ and discuss quantum Hall states for such spaces.
Unlike the case of $S^4$, a finite spatial density can be obtained
with a finite number of internal states for each particle. 
We treat the case of ${\bf C}P^2$ in some detail considering both
Abelian and nonabelian
background fields.
The wavefunctions are obtained and
incompressibility of the Hall states is shown. 
The case of ${\bf C}P^3$ is related
to the case of
$S^4$. 
\end{titlepage}

\newpage
\pagestyle{plain}
\setcounter{page}{2}
\newpage
\noindent
{\bf 1. Introduction}
\vskip .1in
There has appeared recently an interesting extension of the quantum Hall
effect (QHE) to (4+1) dimensions. Hu and Zhang
analyzed the Landau
problem of charged particles moving on
$S^4$ in the background of an $SU(2)$ instanton field \cite{zhang}. 
The particles are in some definite
representation of this $SU(2)$ gauge group. In
considering the many-body problem of
$N$ fermionic such particles, it was found in \cite{zhang} that in order to obtain
a reasonable thermodynamic limit with a finite spatial density of particles, one
has to consider very large
$SU(2)$ representations. Each particle is then endowed with an infinite number of
$SU(2)$ internal degrees of freedom. This feature becomes problematic when one
analyzes the edge excitations.
It is well known that the edge excitations of a
quantum Hall droplet in (2+1) dimensions are described by a massless 
scalar field.
In analogy with this, one expects the edge excitations of the (4+1)-dimensional
droplet to give higher spin massless fields, in particular the graviton, and may
provide an approach to the quantum description of a graviton.  The anlysis in
\cite{zhang} though shows that the spectrum of edge excitations contains massless
particles of all values of spin, rather than just up to spin $2$.
This is related to the infinite $SU(2)$
degrees of freedom.

In this paper, we present another higher dimensional
generalization of the QHE, which avoids this feature of infinite number of internal
degrees of freedom.  We consider charged particles moving on ${\bf C}P^2$, a
topologically nontrivial, compact four-dimensional space, in the background of
$U(1)$ and
$SU(2)$ gauge fields. We show that in the thermodynamic limit, it is possible to
obtain droplet configurations of finite density and finite $SU(2)$ degrees of
freedom. The latter would avoid the problem of arbitrarily high spin edge
excitations. Similar
results are derived for higher even dimensional manifolds ${\bf C}P^k$.
\vskip .2in
\noindent
{\bf 2. QHE on the two-dimensional sphere or ${\bf C}P^1$}
\vskip .1in
We first review the well known case of the QHE on ${\bf C}P^1$ or $S^2$ 
with a constant background magnetic field \cite{haldane}, which can be thought of
as due to a magnetic monopole at the origin if the sphere is embedded in
three-dimensional Euclidean space in the usual way.
Our emphasis is on a group theoretic analysis which can be
easily generalized to higher dimensional cases.

${\bf C}P^1$ can be parametrized by two complex coordinates $u_{\a}$, such that
\be
u^*_{\a} u_{\a} =1 \label{1}
\ee
with the identification $ u_{\a} \sim e^{i \theta} u_{\a}$.
A point $x^i$ on $S^2$ with radius $r$ is written in terms of $u_{\a}$'s as
\be
x^i = r ~u^{\dag} \sigma ^i u \label{2}
\ee
where $\sigma ^i$ are the Pauli matrices.

${\bf C}P^1$ can be mapped to the two-dimensional flat Euclidean space by the standard
stereographical mapping
\be
u= {1 \over {\sqrt{1 + z \bz}}} \left ( \matrix{1  \cr
z  \cr} \right) \label{3}
\ee
where $z={({x+iy})/r}$.

The vector potential on the sphere is 
\be
A = -i~n~ u^*_{\a} d u_{\a} \label{4}
\ee
where $n$ is an integer due to the Dirac quantization rule. Since
\be
\int F = 2 \pi n \label{5}
\ee
where $F=dA$, $n$ is related to the magnetic field $B$ as
\be
n= 2B r^2 \label{6}
\ee

Wavefunctions restricted to ${\bf C}P^1$ with this magnetic field, are such that
$ D \Psi = (\del - i A) \Psi$ is gauge covariant under $u
\rightarrow e^{i \theta} u$. This implies that the wavefunctions have the form
\be
\Psi ( u, u^*)  \sim u_{\a_1}...u_{\a_p} u^*_{\beta_1}...u^*_{\beta_q}
\label{7}
\ee
where $p-q=n$.

$SU(2)$ rotations of $u,~u^{\dag}$ are of the form $u \rightarrow h u,~ u^{\dag}
\rightarrow u^{\dag} h^{-1}$, where $h \in SU(2)$. The generator of this
transformation is 
\be
L^i = {1\over 2}\left[ u_{\beta} \sigma ^i _{\a \b} {\del \over {\del u_{\a}}}-
u^*_{\a} \sigma ^i _{\a \b} {\del \over {\del u^*_{\b}}} \right]\label{8}
\ee
$x^i$ in (\ref{2}) transforms as a vector under this transformation, so $L^i$ are
the angular momentum operators and satisfy the $SU(2)$ algebra
\be
\bigl[ L_i,~L_j \bigr] = -i \e_{ijk} L_k \label{9}
\ee
(The extra minus sign on the r.h.s of (\ref{9}) is consistent with the usual
angular momentum algebra; in making comparisons we identify $u_{\a} = \la \a | u
\ra$.)

In writing down the Hamiltonian, we have to identify the covariant
derivatives on
$S^2$. In order to do this, it is useful to work with the coset representation
\be
{\bf C}P^1=S^2={SU(2) \over U(1)} \label{10}
\ee
This relation shows that functions on $S^2$ can be thought of as functions of
$SU(2)$ which are invariant under the $U(1)$ subgroup. Since a basis of functions
for
$SU(2)$ is given by the Wigner $\cal{D}$-functions, a basis for functions on $S^2$
is given by the
$SU(2)$ Wigner functions
${\cal{D}}^{(j)}_{L_{3}R_{3}} (g)$, with trivial right action of
$U(1)$, in other words, the $U(1)_R$-charge  $R_3 =0$.
In this language, derivatives on $S^2$ can be identified as $SU(2)$
right rotations on $g$ (denoted by $SU(2)_R$) satisfying an $SU(2)$ algebra
\be
\bigl[ R_{+},~R_{-} \bigr] = 2~ R_{3}\label{11}
\ee
where $R_{\pm}= R_1 \pm R_2$. 
$R_{i}$'s are dimensionless quantities. The dimensional covariant derivatives are
\be
D_{\pm} = i~{R_{\pm} \over r}\label{12}
\ee
In the presence of the magnetic monopole, the commutator of the covariant
derivatives is related to the magnetic field, in other words, we need
$[D_+ ,D_-] = -2B$. From (\ref{6}) and (\ref{11}), we see that this fixes
$R_{3}$ to be half the monopole number $n$. Therefore the wavefunctions on $S^2$
with the magnetic field background are of the form
${\cal{D}}^{(j)}_{L_{3}{n \over 2}} (g)$. (The Dirac quantization rule is seen
from this point of view as related to the quantization of angular momentum, as
first noted by Saha \cite{saha}.)

The angular momentum operators $L_{i}$ generate left rotations on $g$. Since the
left and right $SU(2)$-actions on $g$ are independent, the operators $D_i$
and
$L_i$ are mutually commuting
\be
\bigl[ D_i,~L_j \bigr] =0 \label{13}
\ee
The $L_i$ are the magnetic translations on the sphere, commuting with
the covariant derivatives and leading to the degeneracy of the Landau levels.
Since the wavefunctions are of the form ${\cal{D}}^{(j)}_{L_{3}{n \over 2}} (g)$,
with the same $j$-value for left and right actions, we see that
$\sum_{i=1}^3 R_i^2 = \sum_{i=1}^3 L_i^2 = j(j+1)$.

We can now write down the Hamiltonian
\beq
H &=& - {1 \over {4M}} \bigl( D_{+} D_{-} + D_{-} D_{+} \bigr) \nonumber \\
&=& {1 \over {2M r^2}} \bigl( \sum_{i=1}^3 R_{i}^2 - R_3^2 \bigr) \nonumber\\
&=&
{1 \over {2M r^2}} \bigl( \sum_{i=1}^3 L_i^2 - {n^2 \over 4} \bigr)\label{14} 
\eeq
where $M$ is the particle mass.
For the eigenvalue $\half n$ to occur as one of the possible values for 
$R_3$, so that we can form ${\cal{D}}^{(j)}_{L_{3}{n \over 2}} (g)$,
we need $j=\half n + q$, $q=0,1,..$.  
Since $L^2=j(j+1)$,
the energy eigenvalues are
\beq
E_q & = & {1 \over {2M r^2 }} \left[ (\half n + q) (\half n + q +1) - {n^2
\over 4} \right] \nonumber \\
& = & {B \over {2M}} (2 q +1) + {{q(q+1)} \over {2M r^2}}\label{15}
\eeq
The integer $q$ plays the role of the Landau level index. The lowest Landau
level (LLL) or the ground state has energy
$B/2M$ and the states
$q > 0$ are separated by a finite energy gap. The degeneracy of
the $q$-th Landau level
is $2j+1=n+1+2q$.
In the limit $r
\rightarrow \infty$, the planar image of the sphere under the
stereographic map becomes flat and so this corresponds to the standard
planar Landau problem. We see that as $r\rightarrow \infty$,
(\ref{15})
reproduces the known planar result for the energy eigenvalues and the degeneracy. 

In order to derive the wavefunctions, we need the
explicit representation of
$R_{+},~R_{-}$ as differential operators.
For this, one can use the following parametrization
of $g$ in terms of the complex coordinates $u$.
\be
g= \left(\matrix{u_2^* & u_1 \cr
-u_1^* & ~~u_2 \cr} \right) \label{16}
\ee
A $U(1)_R$-rotation leaves the point corresponding to $g$ on ${\bf C}P^1$
invariant, because of the identification $u_{\a} \sim
e^{i\theta} u_{\a}$. The remaining generators of the $SU(2)$ right
rotations are easily identified as
\beq
R_{-} & = &~~\e_{\a \b} u_{\a}^* {\del \over \del u_{\b}} \nonumber \\
R_{+} & = & -\e_{\a \b} u_{\a} {\del \over \del u^*_{\b}} \label{17}
\eeq
Here we are differentiating as if all components of $u_\a$, $u^*_\a$
are independent; the fact that they are not is immaterial since 
$R_i$ preserve the constraint $u^*_\a u_\a =1$.
With the operators given above,
one can easily check that on the wavefunctions (\ref{7})
\be
\bigl[R_{+},~R_{-} \bigr] \Psi = \bigl(u_{\a}{\del \over \del u_{\a}} - 
u^*_{\a}{\del \over \del u^*_{\a}} \bigr) \Psi = n \Psi \label{18}
\ee

We now consider the many-body fermion problem by restricting the fermions to the
LLL level. In this case, $j =\half n$, $R_3 =\half n$, so that
we have the highest weight state for the right action of $SU(2)$.
The LLL condition is thus $R_+ \Psi =0$, with the solution
\be
\Psi_A =\Psi_{{\alpha_1} {\alpha_2}\cdots {\alpha_n}} =  u_{\alpha_1} 
u_{\alpha_2}  \cdots u_{\alpha_n} 
\label{19}
\ee
This has degeneracy $n+1$, so that when the filling fraction is $1$
with all states occupied, we have $N= n+1$ fermions. In (\ref{19}),
$A$ is a composite index taking $n+1$ values.
The many-fermion wavefunction is the Slater determinant given as
\be
\Psi_N = \e^{{A_1} {A_2}\cdots {A_N}}\Psi_{A_1}(u^{(1)})\Psi_{A_2}(u^{(2)})\cdots
\Psi_{A_N}(u^{(N)})
\label{20}
\ee
We write $u^{(i)}_\a$ for the positions of the particles, $i=1,2,\cdots , N$.
The product $\e_{\a\b} u^{(i)}_\a u^{(j)}_\b$ is $SU(2)$ invariant
and antisymmetric under $i\leftrightarrow j$. The Slater determinant 
will involve $N= n+1$ particle labels, the $u^{(i)}$ corresponding to each
particle (for each $i$) should occur $n$ times. Combined with antisymmetry, this
gives
\be
\Psi_{N} ( u^{(i)}_{\a }) \sim \prod _{i<j} \e _{\a\b} u^{(i)}_{\a } u^{(j)}_{\b }
\label{21}
\ee 
In the thermodynamic limit,
$N \rightarrow \infty,~n \rightarrow \infty$, this corresponds to a
configuration of constant density, $\rho = {N /{4\pi r^2}}$ which tends to
${B/{2 \pi}}$ in the planar limit $r\rightarrow \infty$.

The two-point function for the density is also of some interest and can be easily
calculated. If we take $\Psi_N^* \Psi_N$ and integrate over all
particle positions except two, we find
\be
\int d\mu (3, 4,\cdots ,N )~ \Psi_N^* \Psi_N \sim
\vert \Psi^{(1)}\vert^2~\vert\Psi^{(2)}\vert^2~ - ~\vert \Psi_A^{*(1)}
\Psi_A^{(2)}\vert^2
\label{22}
\ee
where $d\mu (3, 4,\cdots ,N )$ is the measure of integration for the
positions of
particles $3, 4, \cdots , N$. $d\mu$ goes like
$dz d\bz /(1+\bz z)^2$ for each particle, but the precise nature of this is
irrelevant for this calculation. With
$z= (x+iy) /r$ and
$n=2Br^2$, we get, as $r\rightarrow \infty$,
\beq
\int d\mu (3, 4,\cdots ,N )~ \Psi_N^* \Psi_N &\sim&
1~-~ \left[ {(1+\bz_1 z_2 ) (1+\bz_2 z_1) \over (1+\bz_1 z_1) (1+\bz_2 z_2
)}\right]^n\nonumber\\ &\approx& 1~-~ \exp\left[ - 2B |\vec{x}_1-\vec{x}_2|^2
\right]\label{23}
\eeq
This shows that the two-point function for the density approaches the 
constant value $1$ at separations large compared to the magnetic length.
The probability of finding two particles at the same point is zero, as
expected.

Result (\ref{23}) is an expression of incompressibility \cite{girvin}. One can
also define Laughlin wavefunctions \cite{laughlin} of fractional filling $\nu =
1/(2l+1)$. They are of the form
\be
\Psi^{(2l+1)}_{N} =  \left( \prod _{i<j} \e _{\a\b} u^{(i)}_{\a } u^{(j)}_{\b }
\right)^{2l+1}
\ee
where $(N-1)(2l+1)=n$. Since $\vec{L}_{tot} = \sum_{k=1}^{N} \vec{L}^{(k)}$
commutes with the factor $\e _{\a\b} u^{(i)}_{\a } u^{(j)}_{\b }$, $\vec{L}_{tot}
\Psi_N^{(2l+1)} =0 $, which implies that the wavefunction is translationally and
rotationally invariant \cite{haldane} and therefore it corresponds to a
configuration of constant density $\rho = {N / {4 \pi r^2}} \rightarrow {B
/ [{2 \pi(2l+1)}]}$.

\vskip .2in
\noindent
{\bf 3. QHE on ${\bf C}P^2$}
\vskip .1in

In this section we shall extend the previous analysis to
${\bf C}P^2$. This space can be parametrized by three complex coordinates
$u_{\a}$, such that
$u^*_{\a} u_{\a} =1$ with the identification $u \sim e^{i \theta} u$.

In analyzing the Landau problem on ${\bf C}P^2$ we use the fact that ${\bf C}P^2$ can be
written as a group coset space
\be
{\bf C}P^2 = {SU(3) \over U(2)} \sim {SU(3) \over {U(1) \times SU(2)}}
\label{24}
\ee
In this case, we can choose background magnetic fields which are $U(1)$
and/or $SU(2)$ gauge fields. We shall first consider the
case of a background $U(1)$ field. Functions on ${\bf C}P^2$ can be obtained from
the Wigner functions for
$SU(3)$, namely,
\be
f(g) \sim {\cal{D}}^{(p,q)}_{L, L_3, Y_L; R, R_3, Y_R} (g)\label{25}
\ee
where $g$ is an $SU(3)$ group element. 
The irreducible representations of $SU(3)$ can be labelled by two integers
$(p,q)$, corresponding to a tensor of the form
$T^{\a_1 \a_2 \cdots \a_q}_{\b_1  \b_2 \cdots \b_p}$
which is symmetric in all upper indices, symmetric in all lower indices and
traceless for any contraction between upper and lower indices.
(Each index takes values $1, 2, 3$.)
$L_i,~R_i$, $i=1,2,3$ are the
left, right
$SU(2)$ generators, with $L_i L_i = L (L+1)$,
$ R_i R_i =R (R+1)$, and $Y$ is the hypercharge operator. As
before, derivatives on
${\bf C}P^2$ can be identified as the $SU(3)_R$-rotations $R_i,~i=4,5,6,7$.
For ordinary functions on ${\bf C}P^2$ we must have singlets
under the $R_i$ of $SU(2)_R$ and also zero charge for the
$U(1)_R$. In other words, $R=0$ and $Y_R =0$.
With a background field along the $U(1)$ direction, the wavefunctions 
are singlets
under the subgroup
$SU(2)_R$ and carry nontrivial
$U(1)_R$ charge. 
This can be expressed as
\beq
R_i & = & 0 ~~~~~~~~i=1,2,3 \nonumber \\
R_8 & = & {{\sqrt{3}} \over 2} Y_R  =  -{{(p-q)} \over {\sqrt{3}}} = -{n \over
{\sqrt{3}}}\label{26}
\eeq
The normalization used for $R_8$ corresponds to
\be
t^8 ={1\over 2\sqrt{3}}\left( \matrix {1&0&0\cr 0&1&0\cr 0&0&-2\cr}\right)
\label{30}
\ee
We have chosen the normalization of the generators as 
$\Tr (t^a t^b) =\half \delta^{ab}$ for the fundamental representation.

The analogue of the
angular momentum operators $L_i,~i=1,..,8$ generate left
rotations, as a result they commute with the derivative operators on ${\bf C}P^2$.

Following the steps that led to (\ref{14}, \ref{15}), we find that the energy
eigenvalues of a charged particle on ${\bf C}P^2$ in the presence of this
background field are
\beq
 E & = & {1 \over {2 M r^2}} \sum_{i=4}^{7} R_i^2 = {1 \over {2 Mr^2}} \left[
\sum_{i=1}^8 R_i^2 - R_8^2 \right] \nonumber \\
& = & {1 \over {2 M r^2}} \left[ C_2 (p,q) -
R_8^2 \right]\label{27}
\eeq
where $r$ is a dimensional scale parameter related to the volume of
${\bf C}P^2$ which is $8 \pi^2 r^4$. $C_2(p,q)$ is the quadratic Casimir of the
$(p,q)$ representation.  It is easily calculated as
\be
C_2(p,q)  =  {1\over 3}\left[{p(p+3)} + {q(q+3)} + {pq}\right]  \label{28}
\ee
The energy
eigenvalues can be written in terms of
$n,q$ as
\be
E = {1 \over {2Mr^2}} \left[  q(q+n+2) + n \right] \label{29}
\ee
The index $q$ plays the role of the Landau level index. The LLL has
energy $n /(2Mr^2)$. States with $q>0$ are separated from the ground state
by a finite energy gap if $n$ scales like $r^2$ as $r \rightarrow \infty$.

The $U(1)$ gauge field can be written as
\be
A = i{2  n \over \sqrt{3}} \Tr \left( t^8 g^{-1} dg \right)
\label{31}
\ee
Here $g$ is an element of $SU(3)$. Evidently, under $g\rightarrow gh$,
$h \in U(2)$, we have $A \rightarrow A - d (n \theta^8 /\sqrt{3} )$.
Thus $dA$ is well defined as a covariant vector or one-form on ${\bf C}P^2$
eventhough $A$ itself is defined only on $SU(3)$.
Since $\Tr (g^{-1} dg )=0$, we can evaluate the trace and write
\be
A = - i~n~ u_\a^* du_\a \label{32}
\ee
where we define $u_\a$ as the element $g_{\a 3}$. It is easily seen that
$u_\a^*
u_\a =1$ and we can parametrize it as
\be
u_\a = {1\over \sqrt {1+\bz \cdot z}} \left(\matrix{1 \cr z_1\cr z_2\cr}\right)
\label{33}
\ee
The field strength corresponding to this potential is
\beq
F &=&  -i~n~ du_\a^* du_\a \nonumber\\
&=&-in \left[ {d\bz_i~ dz_i \over (1+\bz \cdot z)} -{d\bz \cdot z ~\bz\cdot dz
\over (1+\bz \cdot z)^2}\right] \label{34}
\eeq
The field strength is proportional to the K\"ahler two-form on ${\bf C}P^2$; it
corresponds to a uniform magnetic field. ${\bf C}P^2$ has a noncontractible
two-surface and the integral of $F$ over this surface must be quantized following
arguments similar to the Dirac quantization requirement for uniform magnetic
field on a two-dimensional sphere. This requires that $n$ in (\ref{34}) be an
integer, which we have already assumed in calculating the energies. The large $r$
limit can be obtained by writing $z_i = (x_i + i x_{i+2})/r$, $i=1,2$.
The magnetic field $B$ may be defined as $n=2Br^2$ as in the two-sphere case.
The scaling of $n$ as $r^2$ is thus natural from this point of view as well.

The dimension of the $(p,q)$-representation expresses the degeneracy at each Landau
level $q$ and this is given by
\be
dim(p,q) = {{(p+1)~(q+1)~(p+q+2)} \over 2}\label{35}
\ee
We now consider the many-body fermion problem, where all the states of the
LLL level are filled and the filling fraction is $\nu =1$. In this case
$p=n,~q=0~, N=(n+1)(n+2)/2$. In the thermodynamic limit where
$r \rightarrow \infty,~ N \rightarrow \infty$, the density of particles is 
\be
\rho = {N \over {8 \pi^2 r^4}} \rightarrow {{n^2} \over {16 \pi^2 r^4}} = \left({B
\over 2\pi}\right) ^2 = {\rm finite}
\label{36}
\ee
where we have used the fact that 
$vol({\bf C}P^2) = 8 \pi^2 r^4$. 
Unlike the case of the many-body problem on $S^4$ discussed by
Hu and Zhang, this result shows that on ${\bf C}P^2$ with a $U(1)$ background field
the particle density is finite in the thermodynamic limit without the need of
introducing infinite internal degrees of freedom.

We have seen that the potential $A$ has the property
\be
A (g e^{it^8\theta} ) = A - d \left({n\theta \over \sqrt{3}}\right)
\label{37} 
\ee
The wavefunctions must therefore obey the requirement
\be
\Psi (g e^{it^8\theta}) = \Psi (g) ~\exp\left( -{i n \theta /\sqrt{3}}\right)
\label{38}
\ee
For the lowest Landau level the wavefunction is therefore given by
\beq
\Psi_A &\sim& g_{i_1 3} g_{i_2 3}\cdots g_{i_n 3}\nonumber\\
&\sim& u_{i_1} u_{i_2}\cdots u_{i_n}\label{39}
\eeq
Using these one-particle wavefunctions, one can construct the Slater determinant
for the fully occupied ($\nu =1$) state with $N =\half (n+1) (n+2)$
particles. Equations (\ref{20}) and (\ref{22}) hold for this case as well, with
the one-particle wavefunction $\Psi_{A}$ as given above in equation (\ref{39}). The
analogue of the two-point density correlation in (\ref{23}) is also easily
calculated.
\beq
\int d\mu (3, 4,\cdots ,N )~ \Psi_N^* \Psi_N &\sim&
1~-~ \left[ {(1+\bz^{(1)}\cdot z^{(2)} ) (1+\bz^{(2)} \cdot z^{(1)}) \over
(1+\bz^{(1)}\cdot z^{(1)}) (1+\bz^{(2)}\cdot z^{(2)} )}\right]^n\nonumber\\
&\approx& 1~-~
\exp\left[ - 2B |\vec{x}^{(1)}-\vec{x}^{(2)}|^2
\right]\label{40}
\eeq
This may again be taken as the expression of incompressibility.
Further, in the Slater determinant,
the
$SU(3)$ indices of the
$u_{\a}$ are all contracted and hence it is invariant under the
left action of $SU(3)$ which are the magnetic translations.
This leads to uniform density for states of the form
$[\Psi_N]^{(2l+1)}$.

In the case of ${\bf C}P^2$ one can also have a background $SU(2)$ gauge field. In
the presence of both $U(1)$ and $SU(2)$ background gauge fields, the group
theoretical analysis of the single particle eigenstates and the corresponding
energies is somewhat modified. We now label the irreducible representations of
$SU(3)_R$ by $(p+k, q+k')$, corresponding to the tensor
\be
T^{\a_1...\a_q \g_{1}...\g_{k'}}_{\b_1...\b_p \d_{1}...\d_{k}} \equiv
T^{q,k'}_{p,k} \label{41}
\ee
where $p,q$ indicate $U(1)$ indices and $k,k'$ indicate $SU(2)$ indices.

The wavefunctions are of the form
\be
\Psi(g) \sim {\cal{D}}^{(p+k,q+k')}_{L, L_3, Y_L; R, R_3, Y_R} (g) \label{42}
\ee
They carry nontrivial $U(1)_R$ charge and isospin $SU(2)_R$ as specified by
(\ref{41}). This
can be expressed as (we have assumed that $k > k'$)
\beq
R & = & {{k-k'} \over 2},...,{{k+k'} \over 2} \\
R_8 & = & {\sqrt{3} \over 2} Y_R  = {1 \over {2 \sqrt{3}}} \bigl[-2(p-q)+(k-k')
\bigr] = -{n
\over {\sqrt{3}}} \label{43}
\eeq
where $n$ is the $U(1)$ charge. According to our previous discussion $n$ has to be
an integer, which implies that the isospin $R$ takes integer values.

The energy eigenvalues of a charged particle on ${\bf C}P^2$, in the presence of
both $U(1)$ and $SU(2)$ background fields are
\beq
2Mr^2 E & =& C_2(p+k, q+k') - R_8^2 -R(R+1)\nonumber\\
& =& q^2+q(2k-m+n+2) +n(k+1) +k^2+2k+m^2 \nonumber\\
&&\hskip 1.5in -m(k+1) -R(R+1) 
\label{44}
\eeq
where $k'=k-2m$ and $n$ is the $U(1)$ charge given by (\ref{43}).

For a fixed isospin $R$, there correspond several $SU(3)$ representations with
$SU(2)$ indices $k,~k'$ such that
\be
{{k-k'} \over 2} \le R \le{{k+k'} \over 2} \label{45}
\ee
We notice that if $l$ of the upper $k'$ indices are contracted with $l$ of the
lower $k$ indices in (\ref{41}), this corresponds to an $SU(3)$ representation with
increased
$U(1)$ indices $p,q$ in the following way
\be
T^{q,k'}_{p,k} \rightarrow T^{q+l,k'-l}_{p+l,k-l} \label{46}
\ee
This is because of the tracelessness of $T^{q,k'}_{p,k}$.
A state with increased $q$
indicates a higher Landau level state with increased energy. 
Thus if we can form the same value of $R$ from $(k,k')$ and $(k-l, k'-l)$,
the latter will have higher energy. The nonmaximal values of $R$ which one can
form for a given choice of $(k,k')$, namely the values
$\half (k-k'), \half (k-k')+1, \cdots, \half (k+k')-1$, correspond to the higher
Landau levels of a background with lower $(k, k')$-values.
The lowest energy occurs when $R$ corresponds to the maximal value we can get for
a given $(k,k')$.
For this case, we can write
\be
R = {{k+k'} \over 2} = k-m ~~~~~~~~~~~~~~m=0,1,...,{k\over 2} \label{47}
\ee
Substituting this in (\ref{44}) we can express the energy eigenvalues in terms of
$q,~n,~R$ and $m$.
\be
2Mr^2 E = q^2 +q(2R+n+m+2) + n(R+m+1) + (R+m)(m+1) \label{48}
\ee
The lowest energy eigenstates for fixed $n,~R$ (fixed background fields)
correspond to
$q=0,~m=0$. This is the analogue of the LLL condition. In general, we have two
quantum numbers $q$ and $m$ which specify the Landau level.

In order to determine how $n,~R$ should scale in the thermodynamic limit $r
\rightarrow \infty$ so that all the eigenvalues have finite energies and the
energy gap between different Landau levels remains finite, we study two cases.
\vskip .2in
\noindent
A) {\it Pure $SU(2)$ background:} $n=0,~R \ne 0$

\vskip .1in

In order to have finite energy eigenvalues, $R$ should scale
in the thermodynamic limit 
as $R \sim r^2$, upto a dimensional parameter. The number of states for
the LLL is ${\rm dim}(k,k) = {\rm dim }(R,R) = \half {{(R+1) (R+1) (2R+2)}}$.
The corresponding spatial density is 
\be
\rho \sim {{{\rm dim}(R,R)} \over {(2R+1) r^4}} \rightarrow {R^3 \over {2 R r^4}}
\rightarrow {\rm finite} \label{49}
\ee
This case is very similar to the case analyzed by Hu and Zhang. Finite density is
achieved by attaching infinite $SU(2)$ degrees of freedom to each particle.

\vskip .2in
\noindent
B) {\it Combined $U(1)$ and $SU(2)$ backgrounds:} $n \ne 0,~R \ne 0$

\vskip .1in

In this case we can choose either $n$ or $R$ to scale like $r^2$. In particular we
can choose $n \sim r^2$ while $R$ remains finite as $r\rightarrow \infty$.
The number
of states for the LLL is 
${\rm dim}(R+n,R) = \half {{(n+R+1)(R+1)(n+2R+1)} }$. The corresponding spatial
density is 
\be
\rho \sim {{{\rm dim}(R+n,R)} \over {(2R+1) r^4}} \rightarrow {n^2 \over {4 r^4}}
\rightarrow {\rm finite}\label{50}
\ee
The density is finite while each particle has only finite degrees of freedom.
$R=0$ corresponds to a purely $U(1)$ background.
\vskip .2in

We close this section by giving the explicit formula for the $SU(2)$
background field. The generators of $SU(3)$, namely $t^a$, $a=1, ~2, \cdots, ~8$,
can be grouped into the generators of $SU(2)$ given by $t^1, ~t^2, ~t^3$,
the $U(1)$ generator $t^8$ and the coset directions
$t^\alpha$, $\alpha = ~4, ~5, ~6, ~7$. These matrices obey the commutation
rules $[t^a ,t^b ]= if^{abc} t^c$, with structure constants $f^{abc}$.
As we have mentioned before,
${\bf C}P^2$ is a
curved manifold of nontrivial topology, and the metric tensor
on ${\bf C}P^2$
can be written as $g_{ij}= e^\alpha_i e^\alpha_j$, where the tangent frame fields
$e^\alpha_i$ are given by
\be
e^\alpha_i = 2~ i~\Tr ( t^\alpha g^{-1} \partial_i g ) \label{51}
\ee
We choose the background $SU(2)$ gauge field as
\beq
A^i &=& 2~ i~ \Tr (t^i g^{-1} d g) \nonumber\\
F^i_{kl} &=& - {1\over 2} f^{i\alpha\beta} \left( e^\alpha_k e^\beta_l -
e^\alpha_l e^\beta_k\right)\label{52} 
\eeq
In terms of the frame fields $e^\alpha_i$, the field strength tensor
has constant components, given by $f^{i\alpha\beta}$; so (\ref{52}) is what
qualifies as a constant field for ${\bf C}P^2$.
In coupling this to particles, we
use the
$SU(2)$ matrices in the representation corresponding to the Casimir $R (R+1)$. $R$
plays the role of the combination $eB$.
In this language
the
$U(1)$ field strength tensor (\ref{34}) is given, upto the overall factor of $n$,
as
\be
F^8_{kl} = -{1\over 2} f^{8\alpha\beta} \left( e^\alpha_k e^\beta_l -
e^\alpha_l e^\beta_k\right) \label{53}
\ee
The Riemann
curvature tensor of ${\bf C}P^2$ can be calculated as
\be
R^{\alpha\beta}_{kl} = - F^i_{kl}~f^{i\alpha\beta} ~-~ F^8_{kl}~f^{8\alpha\beta}
\label{54}
\ee 
The chosen values of the background field are proportional to the components of
the curvature tensor as well. 
 
\vskip .2in
\noindent
{\bf 4. Generalization to }${\bf C}P^k$
\vskip .1in

We now consider the question of generalizing the QHE to higher even dimensions
$2k$, by considering charged particles moving on ${\bf C}P^k$ in the presence of a
background $U(1)$ gauge field. Since
\be
{\bf C}P^k = {SU(k+1) \over U(k)} \sim {SU(k+1) \over {U(1) \times SU(k)}}
\label{55}
\ee
the
wavefunctions in the presence of a background $U(1)$ magnetic field can be
constructed using the
$SU(k+1)$ Wigner functions ${\cal{D}}^{(p,q)}_{L,R} (g)$ where
$g$ is an $SU(k+1)$ group element. Here $L, R$ stand for two sets of quantum
numbers specifying the eigenvalues of the diagonal generators
for left and right $SU(k)$ actions on $g$ respectively.
The wavefunctions should be singlets under the 
subgroup $SU(k)$ and carry $U(1)$ charges as specified by the
background field. Because of this, we must consider
irreducible representations of $SU(k+1)$ which contain $SU(k)$ singlets.
Such representations can be
labelled by two integers
$(p,q)$, corresponding to a tensor of the form $T^{\a_1
\a_2\cdots \a_q}_{\b_1  \b_2 \cdots\b_p}$ 
which is symmetric in all upper indices,
symmetric in all lower indices and traceless for any contraction between upper and
lower indices as before, where now
$\a ,\b =1,..,k+1$. $L_i,~R_i$, $i=1,..,k^2-1$ are the left, right
$SU(k)$ generators. The fact that the wavefunctions are singlets under $SU(k)_R$
and carry a nontrivial $U(1)_R$ charge is expressed as
\beq
R_i & = & 0, \hskip 1in i=1,..,k^2-1  \nonumber \\
R_{k^2+2k} & = & -\sqrt{k \over {2(k+1)}} ~(p-q) \nonumber\\
&=& - \sqrt{k \over {2(k+1)}} ~n
\label{56}
\eeq
In other words, we may write the wavefunctions as
\be
\Psi \sim {\cal{D}}^{(p,q)}_{L,R} (g) \label{57}
\ee
where indices $R$ are given by (\ref{56}).

The remaining $2k$ generators $R^A$, $A =k^2, k^2+1, \cdots,
(k^2+2k-1)$, play the role of derivative operators. The Hamiltonian is written in
terms of them as
\beq
H & = & {1 \over 2Mr^2} \sum_{A=k^2}^{k^2+2k-1} R_A^2 = {1 \over 2Mr^2} \left[
\sum_{i=1} ^{k^2+2k} R_i^2 -R^2 _{k^2+2k} \right] \nonumber \\
& = & {1 \over 2Mr^2} \left[ C_2(p,q) - {k \over {2(k+1)}} n^2 \right]
\label{58}
\eeq
The quadratic Casimir of the $(p,q)$ representation is
\be
C_2(p,q) = {k \over {2(k+1)}} \left[ p(p+k+1) + q(q+k+1) \right] + {pq \over {k+1}}
\label{59}
\ee
The wavefunctions (\ref{57}) are eigenfunctions of this Hamiltonian
with the energy
eigenvalues
\be
E = {1 \over {2Mr^2}} \left[ q (q+n+k) + \half {n k}   \right]
\label{59a}
\ee
The index $q$ labels the Landau levels. The LLL has energy 
\be
E = {n \over {2Mr^2}} {k \over 2} \sim {k \over 2} B\label{60}
\ee

The left action of the generators commute with the right action and the
Hamiltonian. The left generators are, once again, the magnetic translations and
lead to the
degeneracy of the Landau levels.
The degeneracy for the
$q$-th Landau level
is given by the dimension of the 
$(p,q)$-representation as
\be
dim(p,q) = {{k (p+q+k) (k+p-1)! (k+q-1)!} \over {(k!)^2 p!q!}} \label{61}
\ee
For the LLL we find
\be
dim(n,0) = {{(n+k)!} \over {k! n!}} \label{62}
\ee
The completely filled LLL, $\nu =1$, contains $N$ fermions where $N= dim(n,0)$. In
the thermodynamic limit $r \rightarrow \infty,~ n \rightarrow \infty$, this
corresponds to a configuration of constant density 
\be
\rho \sim {N \over r^{2k}}
\rightarrow {n^k \over {k!r^{2k}}} \sim B^k \label{63}
\ee

The particular case of ${\bf C}P^3$ is worth commenting on.
${\bf C}P^3$ can be viewed as $SU(4)/U(3)$ or as
$SO(5)/U(2)$, as noted recently in \cite{fabinger}. 
(Gauge potentials on this coset space and related properties have also been
analyzed recently in
\cite{JNP}.)
It is also well
known that
${\bf C}P^3$ is locally of the form $S^4 \times S^2$. It is in fact
a nontrivial bundle over $S^4$ with $S^2$ as the fiber. This is the projective
twistor space of Penrose \cite{penrose}. Therefore we can expect that the 
Hu and Zhang results for $S^4$ can be obtained from our discussion by considering
${\bf C}P^3$ and interpreting a part of it, namely $S^2$, as an internal
symmetry. The $U(1)$ background, combined with this $S^2$ will give
the $SU(2)$ background used in \cite{zhang}. 
In comparing our formulae for ${\bf C}P^3$ with \cite{zhang}, 
we must recall that an
$S^2$ has to be factored out to get to $S^4$. From our discussion in section 2, we
notice that the zero-point energy corresponding to $S^2$ is $(1/2Mr^2) ~n/2$.
Removing this, equation (\ref{59a}) becomes, for the case
of ${\bf C}P^3$,
\be
E = {1 \over {2Mr^2}} \left[ q (q+n+3) + n \right]
\ee 
This agrees with \cite{zhang}. 

It is also possible to choose background fields
which are in the Lie algebra of $SU(k)$ (or subalgebras thereof). 
In this case, we do not expect any qualitative changes in the 
main results compared to the analysis of ${\bf C}P^2$.

\vskip .2in
\noindent
{\bf 5. Summary and comments}
\vskip .1in
We studied the Landau problem of nonrelativistic charged particles moving on even
dimensional spaces ${\bf C}P^k$ in the presence of a $U(1)$ background gauge field
and the corresponding lowest Landau level QHE states. 
The number of states per unit volume of ${\bf C}P^k$, in other words the spatial
density of states, is finite as the volume of ${\bf C}P^k$
is scaled to infinity. In the case of ${\bf C}P^2$ we have also considered $SU(2)$
background gauge fields. With combined $U(1)$ and $SU(2)$ backgrounds, it is
possible to choose the particles to be in a fixed finite dimensional
representation of $SU(2)$ and obtain finite spatial density of states as the
volume is scaled to infinity. Thus it is possible to construct quantum Hall
states of fixed uniform density.
For a quantum Hall droplet on a space of the form
$G/H$, the edge states must transform as a representation of the isotropy group
$H$. Our analysis for
${\bf C}P^2$ thus shows that it is possible to obtain edge states  of
finite dimension (or finite number of polarization states), transforming
as any chosen representation of $H= U(2)$, in the
large volume limit. A more detailed study of these edge states is definitely
of interest.

Finally we note that it is possible and interesting to 
extend this kind of analysis
to the noncommutative versions of these spaces 
utilising the framework
of \cite{grosse1, nccp2, chen}.

\vskip .1in\noindent
{\bf Acknowledgements}
\vskip .1in
We thank A.P. Polychronakos for discussions.
This work was
supported in part by NSF grants PHY-9970724, PHY-0070883 and a PSC-32 CUNY award.

\end{document}